\documentclass{aa}
\usepackage{graphicx}
\usepackage{epsfig}
\usepackage{graphics}
\begin{document}%
   \title{Accretion vs colliding wind models for the 
 gamma-ray binary LS~I~+61~303: an assessment}

   \author{G.~E. Romero\inst{1,2,}\thanks{Member of CONICET, Argentina},
	   A.T. Okazaki\inst{3}, 
	   M. Orellana\inst{1,2,}\thanks{Fellow of CONICET, Argentina}
	   \and 
	   S.P. Owocki\inst{4}, 
		     }

   \offprints{Gustavo E. Romero {\em romero@fcaglp.unlp.edu.ar}}
   \titlerunning{Gamma-ray emission from LS~I~+61~303}

\authorrunning{G.E. Romero et al.}

\institute{
Facultad de Ciencias Astron\'omicas y Geof\'{\i}sicas, 
Universidad Nacional de La Plata, Paseo del Bosque, 1900 La Plata, Argentina 
\and 
Instituto Argentino de Radioastronom\'{\i}a, C.C.5, 
(1894) Villa Elisa, Buenos Aires, Argentina 
\and
Faculty of Engineering, Hokkai-Gaukuen University, 
Toyohira-ku, Sapporo 062-8605, Japan 
\and Bartol Research Institute, University of Delaware, 
Newark, DE 19716, USA}

\date{Accepted: 10 August 2007}


\abstract
{LS~I~+61~303 is a puzzling Be/X-ray binary with variable gamma-ray
emission up to TeV energies.  
The nature of the compact object and the origin of the high-energy 
emission are unclear.  
One family of models invokes particle acceleration in shocks from the collision
between the B-star wind and a relativistic pulsar wind,
whereas another centers on a relativistic jet powered by accretion from the
Be star decretion disc onto a black hole.
Recent high-resolution radio observations 
showing a putative ``cometary tail'' pointing away from the Be star
near periastron have been cited as support for the pulsar-wind model.
}
{We wish to carry out a quantitative assessment of these competing models.}
{We apply a ``Smoothed Particle Hydrodynamics'' (SPH) code in
3D dynamical simulations for both the 
pulsar-wind-interaction and accretion-jet models.
The former yields a dynamical description of the shape of the wind-wind
interaction surface.
The latter provides a dynamical estimation of the accretion rate under a
variety of conditions, and how this varies with orbital phase.
}
{
The results allow critical evaluation of how the two distinct
models confront the data in various wavebands.
When one accounts for the 3D dynamical wind interaction under realistic
constraints for the relative strength of the B-star and pulsar winds,
the resulting form of the interaction front does not match the putative
``cometary tail'' claimed from radio observations.
On the other hand, dynamical simulations of the accretion-jet model
indicate that the orbital phase variation of accretion power includes a secondary broad 
peak well away from periastron, thus providing a plausible way to
explain the observed TeV gamma ray emission toward apastron.
}
{
Contrary to previous claims, the colliding-wind model is not clearly established
for LS~I~+61~303, whereas the accretion-jet model can reproduce many key
characteristics, such as required energy budget, lightcurve, and spectrum
of the observed TeV gamma-ray emission.
}


\maketitle

%

\section{Introduction}

Some massive X-ray binaries emit electromagnetic radiation across the
entire spectrum, from radio to TeV gamma-rays. 
Cherenkov gamma-ray telescopes have recently detected variable 
emission from the binary systems PSR B1259-63 (Aharonian et al.  2005a), 
LS 5039 (Aharonian et al.  2005b, 2006a), LS I +61 303 (Albert et al.  2006) 
and Cygnus X-1 (Albert et al. 2007).
The first source is a radio pulsar in a long and eccentric orbit around a Be
star.  
Its high-energy emission is interpreted as inverse Compton (IC)
upscattering of stellar photons by relativistic electrons accelerated
in a colliding wind region (e.g. Tavani \& Arons 1997, Kirk et al
1999, Khangulyan et al. 2007).  
The case of the next two sources is not so clear, since the nature 
of the compact object has not been established\footnote{We note that Casares et al. (2005a) suggest that LS 5039 might be a black hole.} and jet-like radio 
features have been detected (Paredes et al.  2000, Massi et al.  2001).  
Nonetheless, some authors (e.g. Dubus 2006a) have suggested that all 
these sources may be similar to PSR B1259-63, in the sense of being 
powered by colliding winds. 
In contrast, Cygnus X-1 is a well-established microquasar.

In the case of LS I +61 303, recent radio imaging observations along the
orbital period have been cited as evidence that the compact object is an
energetic pulsar around which a synchrotron ``cometary tail" is formed
(Dhawan et al.  2006). Colliding wind models, however, face many problems 
for explaining the spectral energy distribution (SED), 
the time evolution of the emission, the required energy budget, or even the radio morphology.  
Alternative accretion models, wherein the compact object is a
black hole and the non-thermal emission arises from relativistic
jet flows, remain an open possibility (e.g. Bosch-Ramon et al. 2006, Orellana \& Romero 2007).

In this paper we apply a Smoothed Particle Hydrodynamics (SPH) code to
develop more realistic, three-dimensional (3D), dynamical simulations of 
both types of model, and use the results of these simulations to evaluate the
respective merits of each model 
for matching  observational characteristics of LS~I~+61~303.
For the pulsar-wind-collision model a key aim is to compute the 3D
form of the wind-wind interaction front under realistic constraints
for the relative strengths of the B-star and pulsar winds. 
We adopt a non-relativistic pulsar wind model here,
but with a wind momentum adjusted to represent a
relativistic wind with the required total energy.

For accretion models, a key point to establish is the level and phase
variation of accretion power in this system.
Until now, accretion models for LS~I~+61~303 have applied
simplistic estimates based on Bondi-Hoyle 
scaling laws (e.g. Mart\'{\i} \& Paredes 1995,
Gregory \& Neish 2002), ignoring the perturbation of the compact
object on the decretion disc of the Be star, as well as the 
details of the accretion disc formation.  
Our 3D dynamical simulations of the accretion disc include
cases where the compact object is either a neutron star or a black
hole, allowing a quantitative assessment of how well the 
accretion can be used to power jets that might reproduce the observed spectral energy
distribution and its variation with orbital phase.
We can then compare the accretion scenario with the
colliding wind one, discussing the critical issues for each model type.

The structure of the paper is as follows.  In the next section we
review the main known properties of LS I +61 303.  Then we outline the
main features of the accretion/ejection scenario for this source.  

In Sections \ref{sim} and \ref{assess} we respectively describe the simulations for the
 accretion and colliding wind scenarios and present the main results derived from them. Section \ref{discussion} provides a global assessment of both competing scenarios and a brief summary.

\section{The puzzling gamma-ray binary LS~I~+61~303}

LS I+61 303 is a radio-emitting Be X-ray binary discovered by Gregory
\& Taylor (1978).  The primary star in the system is a B0-B0.5Ve star
(Paredes \& Figueras 1986), with a dense equatorial mass ejection that forms a
circumstellar disc (CD).  The distance to the system is $\sim 2$ kpc
(Frail \& Hjellming 1991, Steele et al.  1998).  
The orbital period is well known from radio observations:
$P=26.4960$ days (Gregory 2002).  
The orbital parameters have been determined by Casares et al. (2005b) 
and Grundstrom et al. (2007).  
The orbit is inferred to be quite eccentric,
with eccentricity estimates from $e=0.72\pm0.15$ (Casares et al. 2005b) to
$e=0.55\pm0.05$ (Grundstrom et al.  2007).
Inferred masses are $\sim 12$ $M_{\odot}$
for the primary and $\sim 2.5$ $M_{\odot}$ for the compact object,
assuming an inclination angle $i=30^{\circ}$, but lower masses for the
secondary are possible for larger inclinations (e.g. Dubus 2006a).
The luminosity of the Be star is estimated at
$L_{\star}\sim 10^{38}$ erg s$^{-1}$ (Hutchings \& Crampton 1981). 

Soon after its discovery, LS~I+61~303 was associated with gamma-ray
detections, first with the COS-B source 2EG 135+01 (Gregory \& Taylor
1978), and later with the EGRET source 3EG J0241+6103 (Kniffen et
al.  1997).  
In 2006 
it was detected by the MAGIC telescope as a variable 
gamma-ray source at high energies ($E>200$ GeV, see Albert et al.  2006).  
The maximum in the gamma-ray light curve occurs not during 
periastron passage (at phase $\phi=0.23$), but at phase $\phi=0.5-0.6$.  
The radio emission is stronger between phases $\sim 0.45-0.9$.

In X-rays the source has been observed by a number of instruments
including ROSAT, RXTE, ASCA, Chandra, XMM-Newton, and INTEGRAL 
(e.g. Taylor et al. 1996, Leahy et a. 1997, Greiner
\& Rau 2001, Sidoli et al.  2006, Chernyakova et al. 2006, 
Paredes et al. 2007).  
The X-ray flux is variable, changing from high to low states in a few
hours.  
The X-ray luminosity, however, is never very high, varying
in the range $10^{33}-10^{34}$ erg s$^{-1}$.  
Sidoli et al.  (2006) have compiled the whole SED of the source 
(see their Figure 7). No thermal feature is observed at X-rays.  
The peak of non-thermal emission is in gamma-rays, reaching a luminosity of 
$\sim 10^{35}$ erg s$^{-1}$ in EGRET's energy range. 

At $E>200$ GeV the luminosity is $\sim 10^{34}$ erg s$^{-1}$.  
All this shows that extremely energetic particles are present 
and that a very large energy budget is required to power the source, 
i.e. $L>10^{36}-10^{37}$ erg s$^{-1}$.

\section{The accretion/ejection scenario for LS~I~+61~303}

The idea that the non-thermal emission from LS~I~+61~303 is powered by
accretion onto a compact object was first proposed by Taylor \&
Gregory (1982, 1984). 
When Massi et al.  (2001, 2004) detected jet-like radio features extending 
to $\sim 400$ AU from the core, the source was classified as a microquasar (MQ). 
Since the mass of the compact object is not well-constrained, several authors 
have proposed that the secondary could be a low mass ($\sim 2.5$ $M_{\odot}$)
black hole (Punsly 1999, Massi 2004, Bosch-Ramon et al.  2006, 
Orellana \& Romero 2007).  
The scenario of an accreting pulsar seems to be ruled out by
the low-level of the X-ray 
luminosity\footnote{In the case of an accreting pulsar, 
the infalling matter should impact onto the solid surface, producing 
strong and hard X-ray emission, 
which
is not observed.}, 
unless most of the accreting
matter is ejected by the pulsar (Zamanov 1995, Romero et al. 2005).

If the power engine of LS~I~+61~303 is accretion onto a black hole,
collimated outflows of relativistic particles can be expected, as in other
microquasars and in extragalactic quasars.  
The usual assumption of accreting models is that the accretion rate is 
coupled to the jet power: $L_{\rm jet}=q_{\rm jet} \dot{M}_{\rm accr} c^2$,
where $q_{\rm jet}<1$ and could be as high as 0.1.  
The accretion/ejection models can be divided in two groups according to 
how the high-energy emission originates:
leptonic models (e.g. Bosch-Ramon et al.  2006, 
Bednarek 2006, Gupta \& B\"ottcher 2006), wherein the gamma-rays are the
result of IC interactions of relativistic electrons in the outflow
with photons from the star, the CD or locally produced by synchrotron mechanism; 
and hadronic models (e.g. Romero et al.  2003, 2005; Orellana \& Romero 2007), 
wherein the emission at $E>1$ GeV is dominated by gamma rays from the decay 
of neutral pions that originate in inelastic $pp$ interactions.  
Both types of models predict variability over the orbital period, 
since one expects orbital phase variation in both the power of the jet, 
as well as in the target  density for the relativistic particles 
(either photons or ambient matter).  
An additional source of variability is the absorption of gamma-rays
in the photon fields of the primary star and the CD (e.g. Bosch-Ramon
et al.  2006), which
can lead to the formation of electromagnetic cascades and the suppression
of the gamma-ray signal during periastron passage (Orellana \& Romero 2007).

It may be possible in principle to differentiate between leptonic and
hadronic models through the cutoff of the high-energy emission, since
electrons cool very fast through IC losses in the ambient photon
fields, so their highest energy can hardly go beyond a few TeV. This is
not the case for protons, which can reach higher energies. 
In addition, neutrino emission might be detectable from this type of sources if the
bulk of the gamma-ray emission is of hadronic origin (e.g. Christiansen et
al. 2006, Aharonian et al.  2006b).

An obvious requirement of the accretion/ejection models for the
gamma-ray production in LS~I~+61~303 is that the accretion rate must
be high enough 
to sustain the gamma-ray production with
luminosities of $\sim 10^{34}$ erg s$^{-1}$ at phase $\phi=0.5-0.6$,
where MAGIC detected the source.  
We 
note
that the detected EGRET flux, which is probably due to LS~I~+61~303, 
imposes even more severe constraints ($L_{\gamma} \ga 10^{35}$ erg s$^{-1}$). 
There might be some contribution from background sources, but most of the flux 
should come from the binary system, since the emission is variable (e.g. Torres et al. 2001,  Massi 2004). 
We will discuss this further in Section \ref{sim}.

\section{Numerical simulations of the accretion disc formation around
the compact object in LS~I~+61 303}\label{sim}

\subsection{The code}

Simulations presented here were performed with a three dimensional, smoothed
particle hydrodynamics (SPH) code. The code is basically the same as that used
by Okazaki et al. (2002) and Hayasaki \& Okazaki (2004, 2005, 2006). 
It is based on a version originally developed by Benz (Benz 1990; Benz et al. 1990) 
and then by Bate and his collaborators (Bate, Bonnell \& Price 1995). 
Using a variable smoothing length, the SPH equations with a standard 
cubic-spline kernel are integrated with an individual time step for each particle. 
In our code, the Be disc and the accretion disc are modeled by an ensemble of 
gas particles with negligible self-gravity, while the Be star and the compact object 
are represented by sink particles with the appropriate gravitational mass.  
Gas particles that fall within a specified accretion radius are accreted by the sink
particle.

In order to allow the conversion of the kinetic energy to heat due to viscosity 
and to deal with shocks, our code uses the standard form of 
artificial viscosity with two free parameters $\alpha_{\rm SPH}$ and $\beta_{\rm SPH}$,
which respectively control the strength of the shear and bulk viscosity components 
and that of a second-order, von Neumann-Richtmyer-type viscosity
(Richtmyer \& Morton 1994).

\subsection{The simulations}

To optimize the resolution and computational efficiency of our
simulations, we elected to break the decretion and accretion portions 
of the computation into two separate, but linked parts. 
The first focuses on the decretion of the Be disc, starting from the stellar
equatorial surface radius $R_{\star}$, then following its evolution under the
combined gravitational influence of the B-star and compact object.
Because of the strong photoionization heating from the stellar radiation, 
this Be disc is taken to be isothermal at $0.6 T_{\star}$ 
(Carciofi \& Bjorkman 2006), 
where $T_{\star}$ is the effective temperature of the Be star.
The outward viscous diffusion of material in this Be disc provides an
effective source for capture onto the accretion disc around the compact object, 
taken to occur whenever material reaches
a variable capture radius set by $0.9 r_{\rm L}$, where $r_{\rm L}$ is the
Roche-lobe radius for a circular binary
given approximately by (e.g., Paczynski 1971, Warner 1995)
\begin{equation}
   r_{\rm L} \simeq 0.462 \left( \frac{q}{1+q} \right)^{1/3} D \, .
   \label{eq:roche}
\end{equation}
Here the mass ratio $q=M_{X}/M_{\star}$, where
$M_{X}$ and $M_{\star}$ are respectively the masses of the compact object and Be star, and 
$D$ is the instantaneous distance between the stars. 

Following the  approach adopted by Hayasaki \& Okazaki (2004),
the second part of the simulation focuses on the accretion disc around
the compact object, with the above phase-dependent mass transfer from the 
Be disc simulation now used as an outer boundary condition for 
the accretion process.
The inner edge of the accretion disc is set to a small fraction of the
semi-major axis, $2.5 \times10^{-3} a$,
below which the further accretion onto the compact object is expected 
to occur on rapid time scale that is much shorter than that for the simulated
region. 
The value of the semi-major axis is $a=6.4\times10^{12}$ cm.
Since the role of external heating is less clear than for the Be decretion
disc, the material in this accretion disc is assumed to follow a polytropic 
relation with index $\gamma= 1.2$, 
representing a compromise between the isothermal ($\gamma = 1$) and 
adiabatic limits ($\gamma = 5/3$).

The numerical viscosity is adjusted to keep the Shakura-Sunyaev
viscosity parameter $\alpha_{\rm SS}$ constant, using the approximate relation
$\alpha_{\rm SS} = 0.1 \; \alpha_{\rm SPH}\; h/H$ and $\beta_{\rm SPH}=0$, where
$h$ and $H$ are respectively the smoothing length of individual particles
and the scale-height of the 
decretion/accretion disc
(for additional details, see Okazaki et al.~2002). 
In the Be decretion disc, this scale is related to the distance $r$ 
from the Be star by:
\begin{equation}
 \frac{H}{r} = \left(\frac{c_{\rm s}}{v_{\rm crit}}\right)\left(\frac{r}{R_{\star}}\right)^{1/2},
\end{equation}
where $c_{\rm s}$ is the isothermal sound speed and 
$v_{\rm crit}=(GM_{\star}/R_{\star})^{1/2}$ is 
the critical velocity of the Be star.

We set the binary orbit in the $x$-$y$ plane
with the major axis along the $x$-axis. 
In Be disc simulations, the mass ejection mechanism from
the Be star is modeled by constant injection of gas particles at a radius
just outside the equatorial surface. 
For the primary, we assume a B0V star of $M_{\star} = 12 M_{\odot}$, 
$R_{\star} = 10 R_{\odot}$, and $T_{\star} =
26,000\,{\rm K}$. 
The scale-height of the CD is $\sim 0.03R_{\star}$ 
at $r=R_{\star}$, then increasing outward as $r^{3/2}$. 
For the companion, we take $M_X=2.5M_{\odot}$.
The base density of the Be disc is taken to be
$5 \times 10^{-11}\,{\rm g\,cm}^{-3}$. 
Note
that the Be star has a two-component extended atmosphere: a polar, low-density and 
fast wind ($\sim 10^3$ km s$^{-1}$) and a cool, slow CD. 
The matter supplied from the equatorial stellar surface drifts outwards
because of viscous diffusion. 
The velocity of the CD outflow, within $\sim 10 R_{\star}$, is less than a few km s$^{-1}$. 

For the binary, we adopt an orbital period  $P_{\rm orb}=26.496\,{\rm d}$ 
and an eccentricity $e=0.72$ 
(Casares et al. 2005b).

\subsection{Results}

\begin{figure}[ht!]
\resizebox{\hsize}{!}{\includegraphics{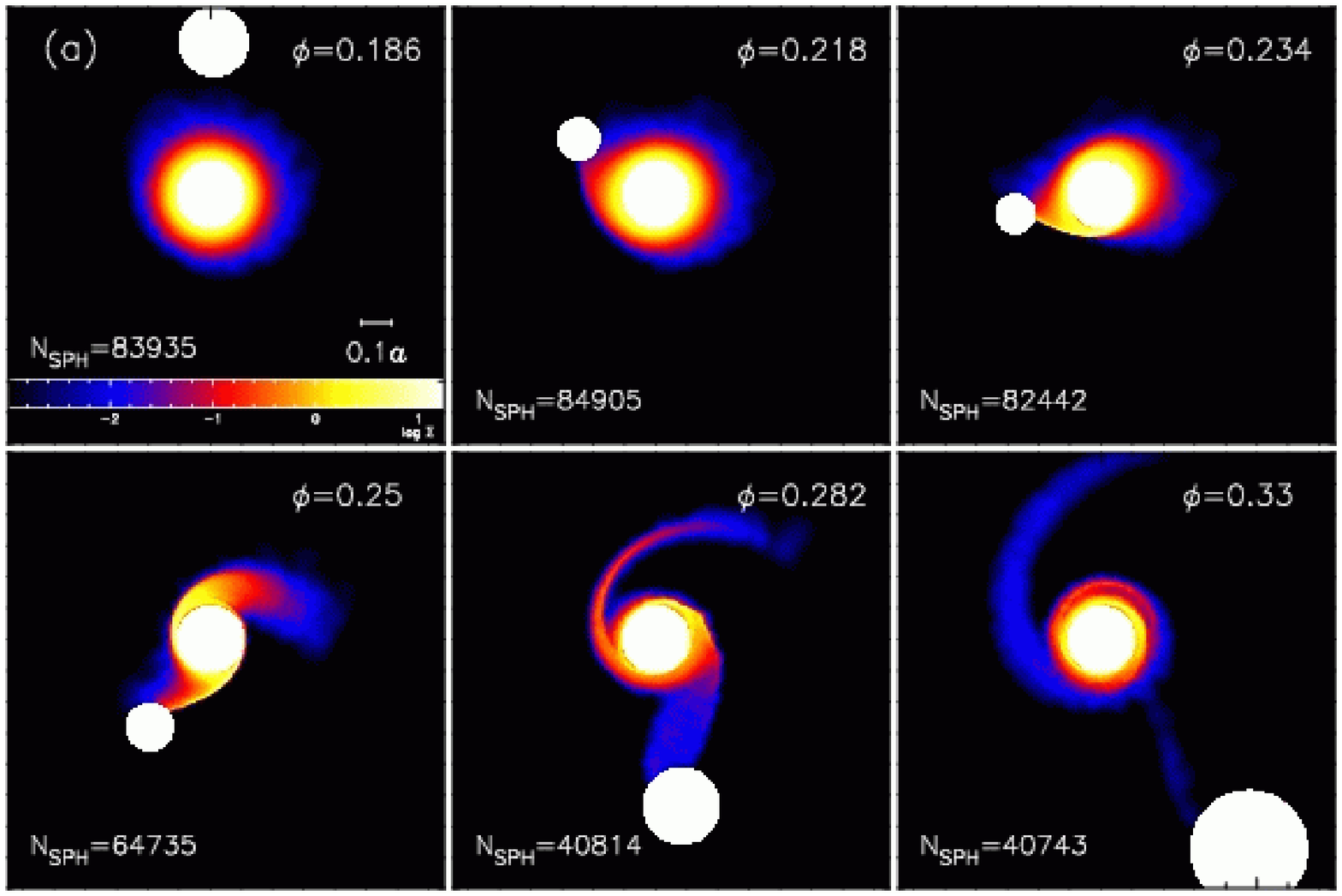}}
\resizebox{\hsize}{!}{\includegraphics{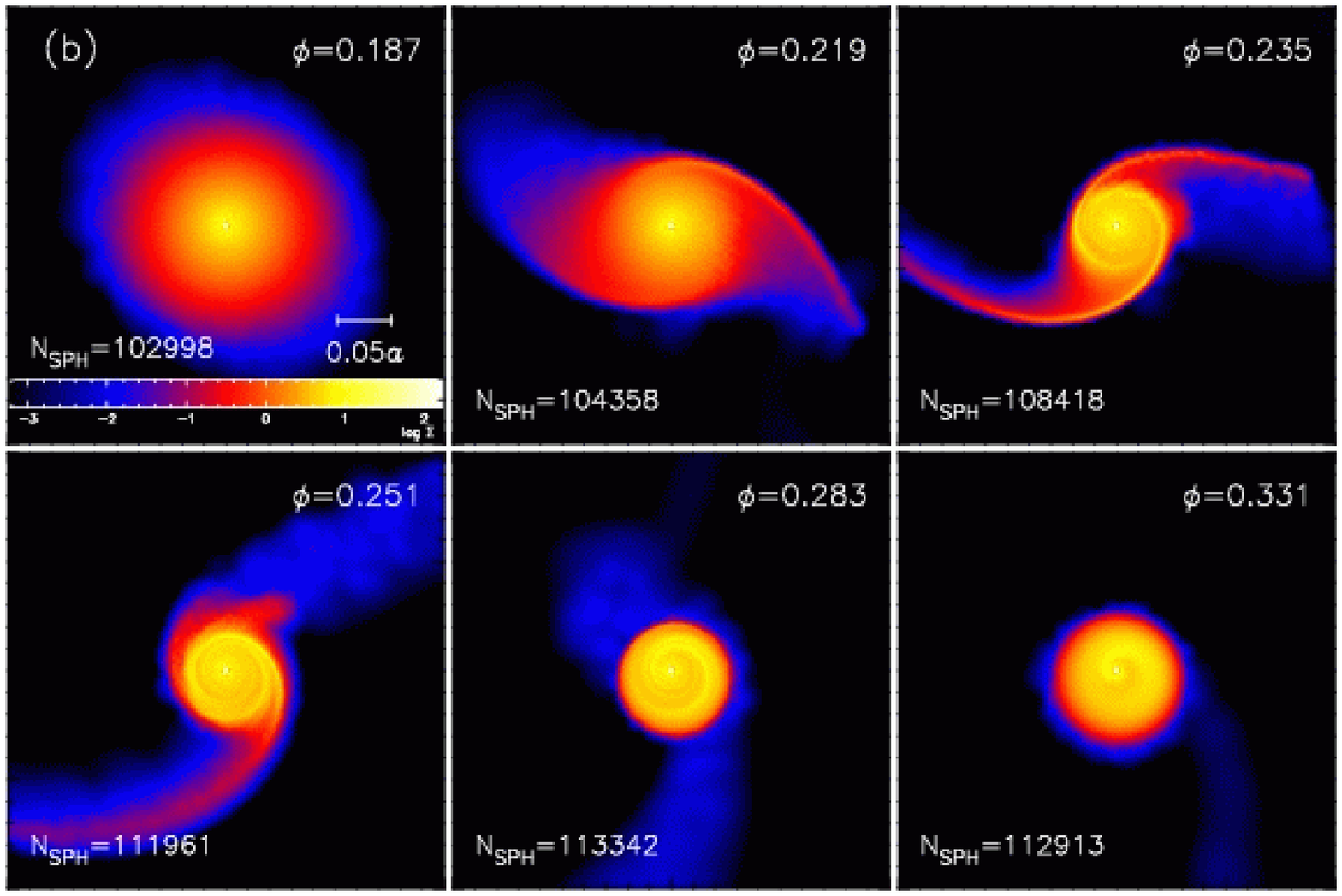}}
\resizebox{\hsize}{!}{\includegraphics{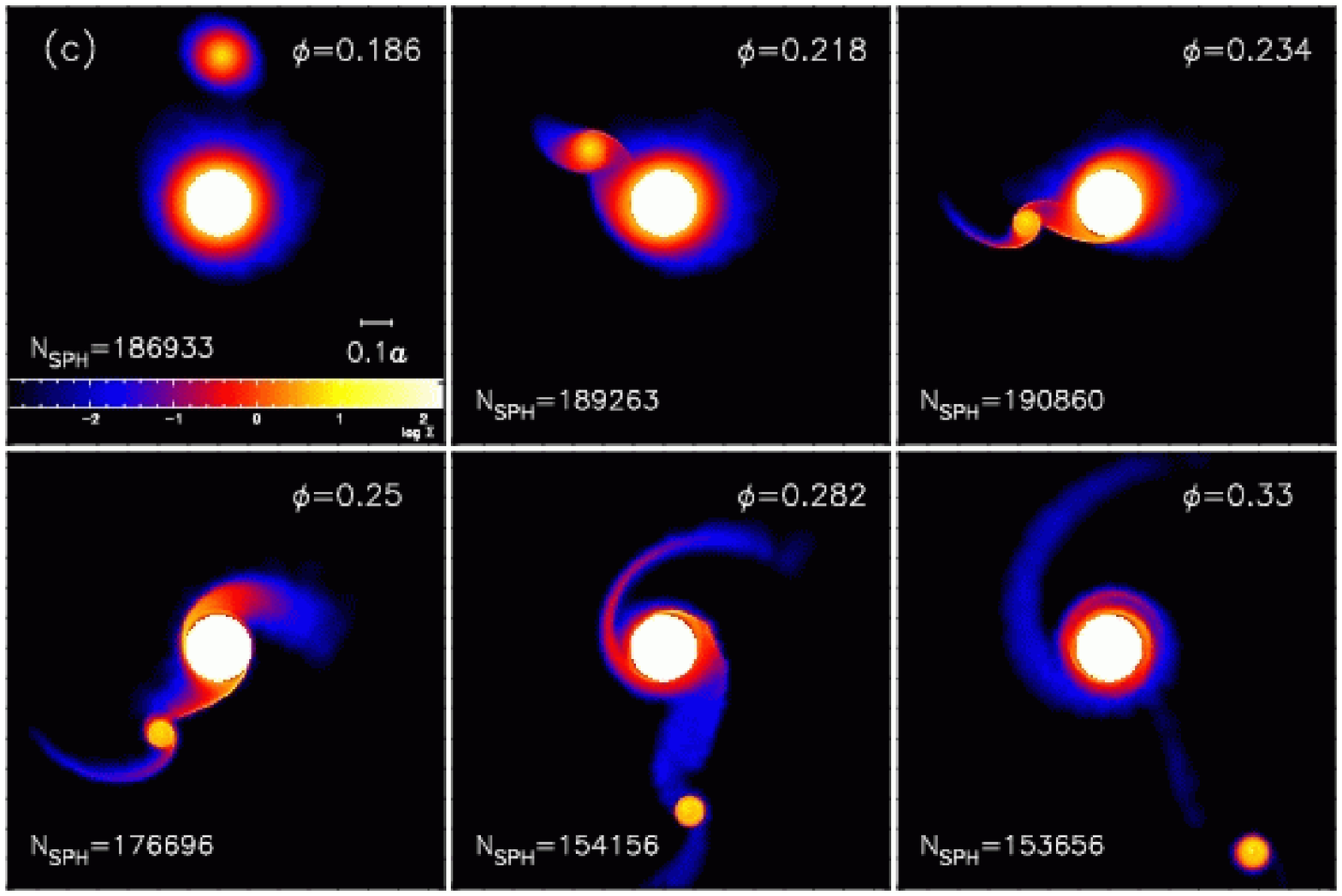}}
\caption{Snapshots from (a) a Be disc simulation and (b) the corresponding 
     accretion disc simulation and (c) their mosaic, covering 
     $\sim 0.15\,P_{\rm orb}$ around 
     periastron passage at $\phi=0.23$. 
     Each panel shows the surface density in cgs units
     on a logarithmic scale. 
     The annotations in each panel give
     the orbital phase and total number of SPH particles. 
     Indicated in the first panel are the size and 
     surface density scales.}
\label{fig:snapshots}
\end{figure}

Let us now examine results from a set of simulations with the Be disc plane
set coplanar with the binary orbital plane. 
Figure~\ref{fig:snapshots} shows snapshots around periastron passage, 
which takes place at the orbital phase $\phi=0.23$, from (a) the Be disc simulation 
and (b) the corresponding accretion disc simulation. 
These snapshots are taken after the Be/accretion disc begins to
show a regular orbital modulation. 
Each panel shows the surface density over four orders of magnitude,
with a logarithmic scaling for the greyscale. 
In Fig.\ref{fig:snapshots}a, where the centre of mass of the system is fixed,
the dark spot near the centre is the Be star, 
whereas the other dark spot represents the compact object with 
the variable accretion radius. 
In Fig.\ref{fig:snapshots}b, the dark spot at the centre is the compact object 
with the fixed accretion radius of $2.5 \times 10^{-3}a$.

Fig.\ref{fig:snapshots} illustrates the interaction between the Be disc and 
the compact object. 
In this highly eccentric system, strong interaction occurs only for a very
short span of time around  periastron passage. 
At periastron, the compact object tidally deforms 
the Be disc and captures particles in the outermost part 
(Fig.\ref{fig:snapshots}a). 
Matter infalling from the Be disc excites the $m=1$ density 
wave in the accretion disc (Fig.\ref{fig:snapshots}b).
At the same time, the accretion disc  as a whole shrinks by the tidal torque 
from the Be star. 
In later phases, the density wave propagates inward, enhancing the accretion rate,
as found by Hayasaki \& Okazaki (2005).
The accretion disc gradually expands until the next periastron passage
of the Be star.

\begin{figure}
\centering\includegraphics[width=0.8\hsize]{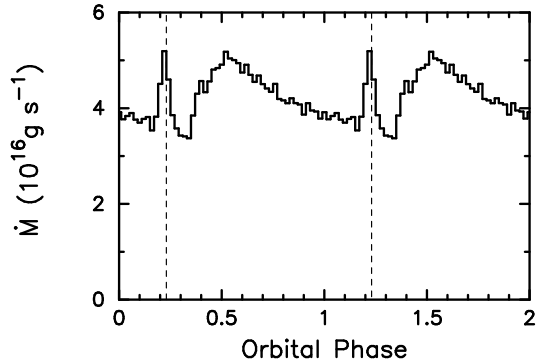}
\caption{Phase dependence of the accretion rate onto the compact object. 
   To reduce the fluctuation noise, the data is averaged over
     three orbital period. 
     The vertical dashed lines denote periastron passage at phase 0.23.}
\label{fig:mdot}
\end{figure}

Figure~\ref{fig:mdot} shows the phase dependence of the accretion rate 
onto the compact object, which is averaged over three orbital periods
to reduce the fluctuation noise. 
Note that this accretion rate has {\em two} peaks per orbit, i.e., a narrow peak
at periastron, and a broad peak at a later phase. 
The periastron peak ($\dot{M} \sim 5.2 \times 10^{16}\,{\rm g\,s}^{-1}$) is due 
to shrinking of the accretion disc. 
In our simulations, particles crossing the accretion radius of the compact object
are taken as having accreted  and so are removed from the simulation. 
Since the accretion disc in the real system should have a much smaller inner radius, 
this peak could be artificial. 
To test this, we carried out a simulation with  the 
accretion radius doubled, to $5 \times 10^{-3}a$.
We find that the periastron peak is indeed higher, 
about five times the value of the standard case,
but the secondary, broad peak has a similar amplitude in both simulations. 
As such, it seems that periastron peak amplitude may be an artifact of the
limited resolution, but the extension over a broad secondary peak
seems more robust.

This secondary peak has value of  $\dot{M} \sim 5.2 \times 10^{16}\,{\rm g\,s}^{-1}$ 
and occurs at  a phase  $\sim 0.52$ that lags the periastron passage by about 0.3
in phase.
As noted above, this secondary peak stems from the inward-propagating, 
$m=1$ density wave in the accretion disc, which is excited by the ram pressure of 
material transferred from the Be disc at periastron.
The phase lag of $\sim 0.3$ results from a combination of
the time for wave excitation and the wave propagation time 
(for a more detailed discussion, see Hayasaki \& Okazaki 2005).

These results have important implications for accretion models of
LS~I~+61~303, since they indicate that a peak in non-thermal emission
could occur well past periastron passage, after phase 0.5, much as
observed by MAGIC (Albert et al.  2006).  
Moreover, the energy available is more than adequate to explain the 
observed high-energy emission.
At phase $\sim 0.5$ the accretion power is $\sim 4.7 \times 10^{37}$ erg s$^{-1}$.  
Assuming that 10\% of this power goes into an outflow -- with just 10\% of that 
converted to relativistic protons --, then a gamma-ray
luminosity of roughly $\sim 8 \times 10^{34}$ erg s$^{-1}$ is expected
from inelastic $pp$ interactions with the medium surrounding the
compact object (Orellana \& Romero 2007).  
If the radiative properties
of the jet are dominated by relativistic electrons or
electron/positron pairs, gamma-ray production through IC interactions
could also explain the observed emission with the inferred
accretion rates (Bosch-Ramon et al.  2006).  
Since our simulations
suggest the periastron peak at phase 0.23 may be artificial,
opacity effects do not need to be invoked to explain the near-apastron 
phase of gamma-rays observed by MAGIC. 
However, as discussed in Section \ref{discussion}, in practice such opacity effects 
may indeed be significant,
making it difficult to detect TeV gamma rays emitted near periastron.

\section{Colliding winds and cometary tails}\label{assess}

The colliding wind scenario for LS~I~+61~303 was first proposed by
Maraschi \& Treves (1981), and then revisited by Dubus (2006a).
Recently, additional authors have proposed that an energetic pulsar
with a relativistic wind could power the gamma-ray source (Chernyakova
et al.  2006, Neronov \& Chernyakova 2007).  
The central idea behind these models is that the collision
between the relativistic pulsar wind and the slow wind from the Be star
should result in a stong shock, thus providing a site for particle
acceleration, much as is thought to occur in other massive binaries 
like WR 140 (Eichler \& Usov 1993, Benaglia \& Romero 2003). 
Relativistic electrons undergoing IC interactions with stellar photons 
then produce the observed gamma-ray emission.  
Since these electrons should also produce synchrotron radiation in 
any ambient  magnetic fields, the colliding wind region itself could be 
detectable as an extended radio source, perhaps even with an elongation 
directed away from the Be star
(Mirabel 2006, see Figure 1, right panel, and Dubus 2006a).

Dhawan et al.  (2006) present a series of VLBA
images, spaced 3 days apart, and covering the entire orbit of
LS~I~+61~303. 
The images show extended radio structures, with the phase near
periastron having indeed an apparent elongation away from the primary
star.  
Since no macroscopic relativistic velocities are measured, 
the authors identify this elongation as
``a pulsar wind nebula shaped by the anisotropic environment, not a
jet''.
Citing the known pulsar PSR B1259-63 as example, 
they conclude that LS~I~+61~303 is a Be-pulsar binary, and not a MQ\footnote{Note, however that no extended radio emission have been found so far in PSR B1259-63.}. 

However, there are several crucial issues for this interpretation.
First, it is important to note that, unlike PSR B1259-63,
no pulsation is detected in LS~I~+61~303.
As such the viability of such a pulsar model requires 
either that any 
pulsed signals must be quenched by a sufficient column of matter throughout 
the orbital phase (including at apastron), 
or, alternatively, that there is a special geometry between the magnetic 
and rotational axis of the putative neutron star.

Second, for a reasonable gamma-ray
production efficiency of a few percent,
the gamma-ray luminosity inferred by EGRET, 
$\sim 10^{35}$~erg~s$^{-1}$, requires the total power of the pulsar wind 
to be higher than several times $10^{36}$ erg s$^{-1}$. 
Assuming a power of $10^{36}$ erg s$^{-1}$, Dubus (2006a) cannot reproduce the 
observed SED, as seen through his Figure 8, which shows the EGRET flux 
to be an order of magnitude above the predicted emission.

Finally, this large required pulsar wind power implies also a large
pulsar wind momentum, requiring then an even stronger Be-wind momentum
if it is to produce the elongated, cometary shape inferred in
the radio images.
The key parameter determining the shape of the interaction surface
of colliding winds is the ratio of wind momentum fluxes, given by 
\begin{equation}
   \eta = \frac{\dot{E}_{\rm PSR}}{\dot{M}_{\rm Be} V_{\rm Be} c}
   \label{eq:eta}
\end{equation}
where $V_{\rm Be}$ and $\dot{M}_{\rm Be}$ are the velocity
and mass loss rate of the Be wind, and  
$\dot{E}_{\rm PSR}$ is the power of the pulsar wind.
Taking $V_{\rm Be}=10^{3}\,{\rm km\,s}^{-1}$ and 
$\dot{M}_{\rm Be}=10^{-8} M_{\odot}\,{\rm yr}^{-1}$ as strong wind
parameters for a B0V star, we have 
$\eta \sim 0.53 \dot{E}_{\rm PSR}/10^{36}\,{\rm erg\,s}^{-1}$.
In a simple 2D model that ignores orbital  motion, the interaction surface
approaches a cone with opening half-angle,
\begin{equation}
    \theta = 180^{o} \, {\eta \over 1 + \eta } \, ,
    \label{eq:opang}
\end{equation}
which can be derived from the analysis of Antokhin et al.
(2004).
Note then that for the above strong Be-wind parameters, even the
minimally required pulsar wind with energy
$\dot{E}_{\rm PSR} \sim 10^{36}\,{\rm erg\,s}^{-1}$ 
would force a quite wide interaction cone, 
with half-angle $\theta \approx 62^{o}$.
And for somewhat larger, but still quite modest pulsar wind energies
$\dot{E}_{\rm PSR} > 2 \times 10^{36}\,{\rm erg\,s}^{-1}$,
the pulsar momentum overwhelms the Be-wind, forcing an opening
half-angle $\theta > 90^{o}$, implying that the interaction cone is
actually wrapped  around the Be-star, {\em not} the pulsar.
(For more detailed discussion of the shape of the interaction surface, 
see, e.g., Antokhin et al.~2004).
Thus even in such simplified 2D interaction models, it is problematic to obtain
an interaction surface that is elongated away from the Be-star, as 
inferred in the radio images.

The situation is further complicated if one takes into account the
azimuthal deflection associated with orbital motion, which can be
particularly  significant near periastron in this highly
eccentric system.
To explore how such orbital effects alter the shape of wind
interaction surface, we have carried 3D SPH simulations of the collision 
between a Be stellar wind (with the above standard mass loss rate and
flow speed\footnote{We remind the reader that the relevant Be-wind here is 
the fast stellar wind with spherically symmetric geometry and not the equatorial CD.}) 
and a much faster flow intended to represent the pulsar wind.
Since our SPH simulations cannot directly simulate a relativistic
flow, we have adopted a fast but non-relativitic wind, with a speed
of $10^{4}\,{\rm km \,s}^{-1}$ for the pulsar wind, adjusting the mass-loss rate
so as to give the same momentum as a relativistic flow with the
assumed energy, 
thus yielding a momentum ratio given as above by 
$\eta \sim 0.53 \dot{E}_{\rm PSR}/10^{36}\,{\rm erg\,s}^{-1}$. 
Our approach is designed to account for the relative ram pressure balance 
between the colliding winds, since this is the primary factor in defining the 
shape of the wind-wind interface. However, by assuming a cooled 
post-shock flow we are neglecting the effect of the gas pressure. We are also neglecting the effect of the Be disc, because it is unlikely that a geometrically thin equatorial outflow with very low velocity could produce the broad ``cometary tail'' inferred from the radio maps.
In the future we plan to investigate models with a more complete
energy balance treatment, including  shock heating and
radiative and other relevant cooling processes.

For simplicity, we take both winds to be isothermal and coasting
without any net external forces, assuming in effect that 
gravitational forces are either negligible (i.e. for the pulsar wind) 
or are effectively canceled by radiative driving terms (i.e. for the 
Be star).
As in the Be and accretion disc simulations, we also ignore the
self-gravity of the wind flows.
The artificial viscosity parameters adopted are $\alpha_{\rm SPH}=1$ and 
$\beta_{\rm SPH}=2$, since we deal now with a situation with no shear 
and a lot of compression.
We set the binary orbit in the $x$-$y$ plane and the major axis of the orbit  
along the $x$-axis. The outer simulation boundary is set at $r=5.25a$ from
the centre of mass of the system. This means scales of $\sim 1$ mas as seen 
from the Earth. Particles crossing this boundary are removed
from the simulation.

\begin{figure}
\centering\includegraphics[width=0.8\hsize]{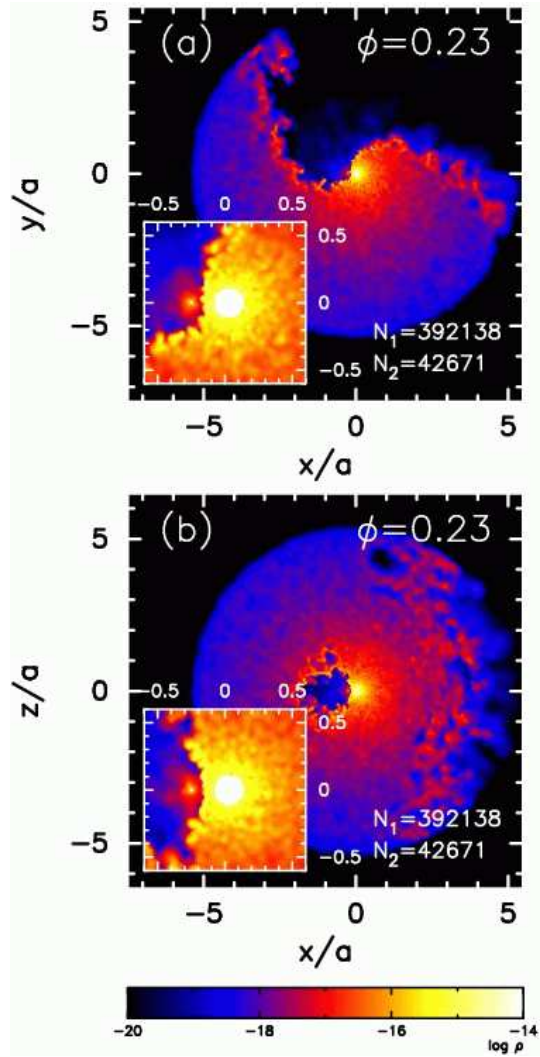}
\caption{
Wind collision interface geometry at periastron for collision between 
a Be-star wind with a pulsar wind of energy 
$\dot{E}_{\rm PSR}=10^{36}\,{\rm erg\,s}^{-1}$, 
corresponding to pulsar/Be wind momentum ratio of $\eta = 0.53$. 
The greyscale plot shows the density in (a) the orbital plane 
and (b) the plane which is perpendicular to the orbital plane and through 
the major axis of the orbit,
on a logarithmic scale with cgs units.
The dark spot near the origin represents the Be star, while
the small dark spot to the left represents the pulsar.
The inset shows the close-up view of the region near the stars.
Annotations in the figure give the orbital phase $\phi$ and the
number of particles in the Be wind, $N_{1}$, and in the pulsar wind, $N_{2}$.
The outer simulation boundary is set at a radius $r=5.25a$ from
the centre of mass of the system.}
\label{fig:cw}
\end{figure}

Figure~\ref{fig:cw} shows a snapshot of a colliding wind model for LS~I~+61~303
with $\dot{E}_{\rm PSR}=10^{36}\,{\rm erg\,s}^{-1}$, and thus $\eta =
0.53$.
On a logarithmic scale with cgs units, the greyscale shows the density 
in (a) the orbital plane and (b) the plane perpendicular to 
this orbital plane and through the orbit's major axis. 
Darker and lighter regions represent
respectively the Be wind and the pulsar wind. 
The dark spot near the origin of the figure represents
the Be star, and a small dark spot on the left side of the Be star is the pulsar.
Note that the density in the pulsar wind is negligible in this simulation.

Although the interaction surface in the simulation exhibits variations from 
instabilities, its global shape is easily traced and illustrates
clearly our key central point.
Namely, when orbital effects are included, even the most favorable assumptions
toward a large Be/pulsar wind momentum ratio do not produce the
simple elongated shape inferred in the VLBI radio image, which was previously
cited as strong evidence in favor of a pulsar wind interaction
scenario.


\section{Discussion and conclusions}\label{discussion}

The previous section shows that the colliding wind model has serious 
problems, both in matching the observed morphology at radio
wavelengths, and in accounting for the observed gamma-ray emission 
through a pulsar  wind with a power of the order of $10^{36}$ erg s$^{-1}$.  
An additional problem is the observed gamma-ray phase variation, which
shows a  peak after phase 0.5.  
Detailed modeling of adiabatic losses of
electrons or very strong Compton losses during periastron passage
(resulting in a high-energy cutoff) could explain a low flux at phase
0.23 (see Khangulyan et al. 2007 for a treatment of this kind in the
case of PSR B1259-63);
but it remains difficult to explain why the gamma-ray 
source is not detected {\sl before} periastron passage, 
as it is in PSR B1259-63. 

Orientation effects might be invoked to explain such a situation, but detailed modeling is lacking.  
In the accretion models, these problems are solved in a natural way by the variable
accretion rate (see Section \ref{sim}).

Opacity effects to gamma-ray propagation in the ambient photon fields
of the Be star and disc offer an additional possibility for quenching 
the high-energy emission close to periastron passage.  
Our SPH simulations provide the azimuthally averaged 
surface density profiles for the Be disc, and together with the
assumption that the disc is isothermal with a temperature 
$T_{\rm disc}=15,600$ K, this allows us to model the emission of
disc photons.
Adding then the photon field of the Be star when accounting for the
finite stellar size (Dubus 2006b), we apply the basic approach of 
Becker \& Kafatos (1995) to compute the photon-photon attenuation of 
gamma-rays generated at the base of
the jet (a height $\sim 10^7$~cm above the black hole),
and directed toward the observer (see Appendix A for further details).
Figure \ref{tau} plots the phase variation of the total opacity along the orbit, 
for two different photon energies.
We see that at periastron the TeV gamma-rays have an optical depth
$\tau \approx 2$, corresponding to a reduction by a factor 
$\exp(-2) \approx 0.14$, implying that longer
integrations will be required to detect the attenuated source.  
But since the absorption is low during the rest of the orbit, 
colliding wind models might predict a detection not only after, 
but also before periastron passage, unless effects related to the viewing angle are invoked.
In any case, the colliding wind model would also face some problems 
to explain the lightcurves at other wavelengths, where the radiation 
peaks at similar phases as the VHE emission.

\begin{figure}
\centering\includegraphics[angle=-90,width=1.1\hsize]{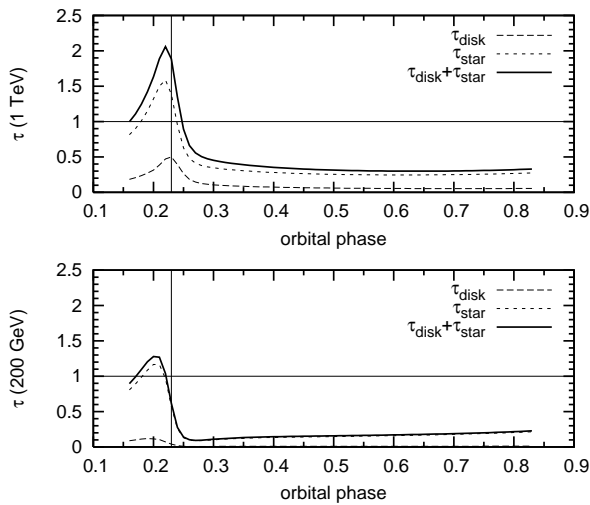}
\caption{Optical depth for the propagation of $\gamma$-ray photons
with energy $E=200$ GeV (lowest energy of MAGIC detection) in the direction to the observer, 
and $E=1$ ~TeV. Circumstellar disc and stellar contributions are shown separately.
The vertical line indicates the phase of periastron passage.}
\label{tau}
\end{figure}

We note that the accretion model is also not free of problems.  
The changing morphology of the radio emission (Dhawan et al.  2006)
requires a highly unstable jet.  
This was already suggested by the Massi et al.  (2004) radio maps.  
The origin of this instability may be related to the density waves 
in the accretion disc found in this paper, but detailed simulations are required to 
reproduce the observed phenomenology. Alternatively, the impact of wind inhomogeneities 
could cause the deflection of the jet (Bosch-Ramon et al. 2006).

Another issue concerning accretion models is the low X-ray luminosity
observed from this system.  
The peak accretion rate, when expressed in Eddington units, 
is $\dot{m}\sim 0.01$.  
Advection-dominated accretion solutions exist for the inner accretion disc 
up to $\dot{m}<\dot{m}_{\rm crit}\sim 0.05-0.1$ (Narayan et al.  1998, see
p.160).  
Since the viscosity parameter and $\dot{m}_{\rm crit}$ are
related through $\dot{m}_{\rm crit}\sim \alpha^2$, 
the models discussed in Section \ref{sim} are consistent with advective,
underluminous accretion, which could explain the paucity of thermal
X-ray emission in LS~I~+61~303.

In addition, since the global change of the accretion rate predicted
by the simulations is $\sim 30$ \% and the opacity is negligible close
to the apastron, accretion-jet models should introduce effects of the
changing environment (photon fields for IC scattering, surrounding
matter for $pp$ interactions) to explain a non-detection at phases
beyond $\sim 0.7$ through longer exposures of current Cherenkov
telescopes. 
In this sense, the study of 
Be-disc mass
not captured by the compact object is quite interesting 
since it could be used to explain the infra-red excess observed 
in the source (Mart\'{\i} \& Paredes 1995) and
provide an additional source of perturbations and interactions for the jet.

Rapid changes in the jet should produce rapid variability in the
gamma-ray emission, that could be observed with GLAST. AGILE and GLAST
observations are crucial to determine the MeV-GeV source luminosity
and spectrum.  MAGIC II could be able to detect the source at the
periastron with significant integrations, yielding information on the
opacity effects and their origin.  Neutrino detection or
non-detection with ICECUBE will shed light on the nature of the
gamma-ray emission.

We conclude that the current evidence favors an accretion/ejection
scenario for LS~I~+61~303.  The debate, however, is not closed and we
can expect to learn new lessons from this fascinating source in the
near future.

\begin{acknowledgements}
We thank an anonymous referee for constructive comments. We also benefited from valuable 
discussions with V. Bosch-Ramon and comments by D. Khangulyan. 
G.E.R. and M.O. research has been
supported by CONICET (PIP 5375) and the Argentine agency ANPCyT through Grant
PICT 03-13291 BID 1728/OC-AR.
A.T.O. thanks the Bartol Research Institute, University
of Delaware, USA for the warm hospitality during his sabbatical visit.
He also acknowledges Japan Society for the Promotion of Science for
the financial support via Grant-in-Aid for Scientific Research
(16540218).
SPH simulations were performed on HITACHI SR11000 at Hokkaido
University Information Initiative Center and SGI Altix~3700 at
Yukawa Institute of Theoretical Physics, Kyoto University.
S.P.O. acknowledges partial support of NSF grant 0507581.

\end{acknowledgements}
{}

\appendix
\section{Calculation of the $\gamma$-ray opacity in the star+disc photon field}
\begin{figure}[ht!]
\resizebox{\hsize}{!}{\includegraphics{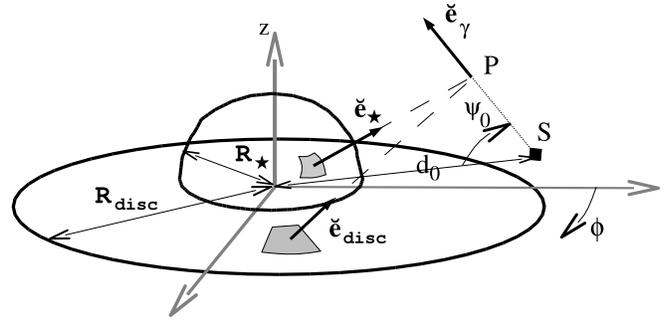}}
\caption{
Scheme of the geometry considered for the propagation of a gamma ray
emitted at $S$ and absorbed at position $P$.  The absorber field is
formed by photons of the stellar surface and its circumstellar disc.}
\label{geo-tau}
\end{figure}

This appendix briefly summarizes the expressions used for the
calculation of the optical depth $\tau(E)$ given in Figure \ref{tau}.
We follow the treatment given by Dubus (2006b) and Becker \& Kafatos
(1995).  The differential absorption probability seen by a
$\gamma$-ray of energy $E$ located at a position $P$ and traveling in
the direction $\mathbf{\breve{e}}_\gamma$ through a photon field with
photons of energy $\epsilon$ has two terms:
\begin{eqnarray}
\mathrm{d}\tau_{\gamma \gamma}&= \mathrm{d}\epsilon
\mathrm{d}l\,[&(1-\mathbf{\breve{e}}_\gamma
\cdot\mathbf{\breve{e}}_\star) n_\star(\epsilon)\,\sigma_{\gamma
\gamma} \mathrm{d}\Omega_ \star +\nonumber\\
&&(1-\mathbf{\breve{e}}_\gamma\cdot \mathbf{\breve{e}}_{\rm
disc})n_{\rm disc}(\epsilon)\, \sigma_{\gamma
\gamma}\mathrm{d}\Omega_{\rm disc}]
\end{eqnarray}
where $\mathbf{\breve{e}}_\gamma$, $\mathbf{\breve{e}}_{\rm disc}$ and
$\mathbf{\breve{e}}_\star$ are unit vectors in the photon direction of
propagation; $\mathrm{d}\Omega$ is the solid angle, seen from $P$, of
the surface emitting the photons (additional angles, not shown in the figure, 
are defined to parametrize each 
surface, star and disc.)
The radiation density, $n(\epsilon)$, is in
photons~cm$^{-3}$~erg$^{-1}$~sr$^{-1}$.  The length $l$ of the
$\gamma$-ray path is measured from its emission point $S$, which is at
a distance $d_0$ and forming an angle $\psi_0$ from the center of the
star (see a sketch in Figure \ref{geo-tau}).  The interaction
cross-section $\sigma_{\gamma \gamma}$ depends only on the square of
the total energy of the two interacting photons in the center of mass frame 
(which, in turn, depends on the collision angle between the photons, 
Gould \& Schr\'eder, 1967):
\begin{equation}
s=\frac{\epsilon E}{2 m_e^2 c^4}(1-\mathbf{\breve{e}}_\gamma\cdot
\mathbf{\breve{e}}_{\star/\rm disc});\;\;
\beta=\left[1-\frac{4(m_e c^2)^2}{s}\right]^{1/2}
\end{equation}
\begin{equation}
\sigma_{\gamma \gamma}=\frac{\pi
r_0^2}{2}(1-\beta^2)\!\left[2\beta(\beta^2-
2)+(3-\beta^4)\ln\left(\frac{1+\beta}{1-\beta}\right) \right]\!,
\end{equation}
where $r_0=e^2/m_e c^2$ is the classical electron radius.  The
integration over $\epsilon$ should be done above the threshold energy
for pair creation, whereas $l$ is integrated up to infinity in order
to compute the probability of detection by a distant observer.

We have considered the star and the circumstellar disc as black-body
emitters\footnote{We note that the emission of outflowing viscous
discs of Be stars could be properly modeled as in Porter (1999).
However, in the present case, in order to account for the luminosity
attributed by Casares et al.  (2005b) to the circumstellar disc
contribution in LSI~+61 303, we have considered a black-body
approximation.}, characterized by $T_\star$ and $R_\star$, and $T_{\rm
disc}$ and $R_{\rm disc}$, respectively.  At distances greater than a
few stellar radii the integration over solid angles reduces to the
point-source approximation.

The trajectory considered for the $\gamma$-ray photons starts at the
orbital position of the compact star, assuming that the high-energy
emission originates close to it.  The orbital variation of the opacity
through the dependence of $d_0$ and $\psi_0$ with the orbital phase
was calculated as in Dubus (2006b).

\end{document}